\documentclass[11pt]{article}

\usepackage{titlesec}
\titleformat{\section}{\large\bfseries}{\thesection}{1em}{}
\usepackage[margin=1.4in]{geometry}

\usepackage{setspace}
\usepackage{amssymb}
\usepackage{amsthm}
\usepackage{amsmath}
\usepackage{amscd}
\usepackage{cite}
\usepackage{epsfig}
\usepackage{epstopdf}
\usepackage[shortlabels]{enumitem}
\setlist[enumerate]{itemsep=0mm}
\usepackage{subfigure}
\usepackage{graphicx}

\usepackage{stmaryrd}
\usepackage{amssymb}
\usepackage{amsmath}
\usepackage{amscd}
\usepackage[ruled,vlined]{algorithm2e}
\usepackage[noautoalign, vcentermath, enableskew]{youngtab}\Yboxdim{12pt}
\usepackage{array}

\date{}
\newtheorem{thm}{Theorem}[section]

\newtheorem{lem}{Lemma}[section]
\newtheorem{prop}{Proposition}[section] 

\theoremstyle{remark}
\theoremstyle{definition}

\newtheorem{defn}{Definition}[section]

\newenvironment{my_enumerate}{
\begin{enumerate}[(1)]
  \setlength{\itemsep}{1pt}
  \setlength{\parskip}{0pt}
  \setlength{\parsep}{0pt}}{\end{enumerate}
}
\usepackage{fancyhdr}
\fancypagestyle{plain}{%
  \fancyhead[L]{\footnotesize {This is a preprint of a paper whose final and definite form has been published in: \\
Discrete Mathematics, Algorithms and Applications Vol. 6, No. 2 (2014) 1450022 (16 pages), \\DOI: 10.1142/S1793830914500220.}}%
}

\title{Priority-based task reassignments in hierarchical 2D mesh-connected systems using tableaux}

\author{Dohan Kim\footnote{E-mail: dkim@airesearch.kr}\\\\
\footnotesize A.I. Research Co., 2537-1 Kyungwon Plaza 201, Sinheung-dong, Sujeong-gu, \\\footnotesize Seongnam-si, Kyunggi-do, 461-811, South Korea\\}

\begin{document}
\maketitle
\begin{abstract}
Task reassignments in 2D mesh-connected systems (2D-MSs) have been researched for several decades. We propose a hierarchical 2D mesh-connected system (2D-HMS) in order to exploit the regular nature of a 2D-MS. In our approach priority-based task assignments and reassignments in a 2D-HMS are represented by tableaux and their algorithms. We show how task relocations for a priority-based task reassignment in a 2D-HMS are reduced to a jeu de taquin slide.
\end{abstract}

\section{Introduction}
A distributed system is a collection of processing nodes connected by an interconnection network~\cite{ramakrishnan1991, tanenbaum1995}. Among various interconnection networks for distributed systems, a two-dimensional (2D) mesh has received extensive study due to its simplicity, efficiency, and structural regularity~\cite{yoo2000, yoo2002, chen2009, shen1997}. Some data structures, such as matrices and arrays, naturally fit into a 2D mesh-connected system (2D-MS)~\cite{shen1997}. Since tasks are often assigned to a submesh in a 2D-MS, continuous submesh allocations and deallocations of different sizes may cause \emph{fragmentation}~\cite{yoo2000, yoo2002} in a 2D-MS. Task relocation is an approach to decrease fragmentation by reassigning a running task to an idle processing node, which involves capturing and transferring the state of the running task to the idle node in a 2D-MS~\cite{yoo2000}. 

Although the construction of a 2D-MS using heterogeneous nodes is considered in~\cite{chen2009}, most of the traditional approaches~\cite{yoo2000,yoo2002,shen1997,seo2005,ababneh2006} are based on the assumptions that nodes in a 2D-MS are homogeneous. Further, the topology of the assigned tasks on a 2D-MS is restricted (e.g., rectangular or square-mesh shape) and priority-based task assignments and reassignments have not often been considered. Therefore, there is a lack of systematic mechanisms of task relocations in a heterogeneous 2D-MS.

For decades, a wide variety of ways to tackle task assignment and reassignment problems in a distributed system have been researched, such as graph-theoretic~\cite{kafil1998, shen1985, wang1988, manneback1994, bokhari1979}, mathematical programming~\cite{chu1980}, and heuristics~\cite{efe1982}. One of the common methods to represent and solve a task assignment problem is a graph-theoretic method using a graph-matching algorithm~\cite{kafil1998, shen1985, wang1988}. To complement the graph-theoretic method, our previous work~\cite{kim2010} presented a \emph{Young tableaux}~\cite{sagan2001, fulton1997, stanley1997, zhao2008} approach to representing task assignments.  

We use tableaux and their algorithms for priority-based task reassignments in a hierarchical 2D-MS, where a hierarchical 2D-MS (2D-HMS) is defined as a 2D-MS consisting of heterogeneous nodes whose priorities (or execution rates) of rows and columns are sorted in descending order. In this paper we convert a 2D-HMS into a \emph{Young diagram}~\cite{sagan2001, stanley1997} and represent a task assignment of a 2D-HMS using a tableau. Our greedy task relocation policy is based on a 2D-HMS in which task relocations are performed systematically by using tableau algorithms.

The remainder of this paper is organized as follows. We provide an introduction to tableaux and their algorithms in Section~\ref{sec:YoungTableaux}. Section~\ref{sec:MeshGraphsAndYoungTableaux} presents a representation of a 2D-HMS using a Young diagram. In this section we define a hierarchical 2D mesh tableau in order to represent a priority-based task assignment and its reassignments in a 2D-HMS. Section~\ref{sec:Priority-basedTaskReassignmentsAndYoungTableaux} shows how task relocations in a hierarchical 2D mesh tableau under the greedy task relocation policy are reduced to a jeu de taquin slide, and how they are applied to a 2D-HMS. Finally, we conclude in Section~\ref{sec:ConclusionsOfPrioritybasedTaskAssignment}.

\section{Preliminaries}
\newcommand{\nten}{\mbox{10}}
\newcommand{\neleven}{\mbox{11}}
\newcommand{\ntwelve}{\mbox{12}}
\newcommand{\nthirteen}{\mbox{13}}
\newcommand{\nfourteen}{\mbox{14}}
\newcommand{\nfifteen}{\mbox{15}}
\newcommand{\nsixteen}{\mbox{16}}
\newcommand{\bfive}{\bf{5}}
\newcommand{\beight}{\bf{8}}
\newcommand{\bnine}{\bf{9}}
\newcommand{\bten}{\bf{10}}
\newcommand{\pone}{{\textrm{ }1}^{{ }^{{}^{\textrm{ }\textrm{ }1}}}}
\newcommand{\ptwo}{{\textrm{ }2}^{{ }^{{}^{\textrm{ }\textrm{ }2}}}}
\newcommand{\pthree}{{\textrm{ }3}^{{ }^{{}^{\textrm{ }\textrm{ }3}}}}
\newcommand{\pfour}{{\textrm{ }4}^{{ }^{{}^{\textrm{ }\textrm{ }4}}}}
\newcommand{\pfive}{{\textrm{ }5}^{{ }^{{}^{\textrm{ }\textrm{ }5}}}}
\newcommand{\psix}{{\textrm{ }6}^{{ }^{{}^{\textrm{ }\textrm{ }6}}}}
\newcommand{\pseven}{{\textrm{ }7}^{{ }^{{}^{\textrm{ }\textrm{ }7}}}}
\newcommand{\peight}{{\textrm{ }8}^{{ }^{{}^{\textrm{ }\textrm{ }8}}}}
\newcommand{\pnine}{{\textrm{ }9}^{{ }^{{}^{\textrm{ }\textrm{ }9}}}}
\newcommand{\pten}{{10}^{{ }^{{}^{10}}}}
\newcommand{\peleven}{{11}^{{ }^{{}^{11}}}}
\newcommand{\ptwelve}{{12}^{{ }^{{}^{12}}}}
\newcommand{\pthirteen}{{13}^{{ }^{{}^{13}}}}
\newcommand{\pfourteen}{{14}^{{ }^{{}^{14}}}}
\newcommand{\pfifteen}{{15}^{{ }^{{}^{15}}}}
\newcommand{\psixteen}{{16}^{{ }^{{}^{16}}}}
\label{sec:YoungTableaux}
This section provides necessary definitions and terminology used in this paper. Definitions and results in this section are found in~\cite{chiu1999,yoo2000,fulton1997,hungerford1980,sagan2001,stanley1997,zhao2008,kim2010,vessenes2004,fraleigh1998,frame1954,shi2006,sinnen2007,suter2004,seo2005,ababneh2006}.

A \emph{heterogeneous system} $N$ is a set of heterogeneous nodes $N=\{n_1, n_2,\ldots, n_m\}$ whose communications are described by a network topology. By a \emph{node} we mean a processor (or agent) that carries out a task. A heterogeneous system $N$ is said to be \emph{consistent} if node $n_a \in N$ executes a task $d$ times faster than node $n_b \in N$, then it executes all other tasks $d$ times faster than node $n_b$. In a consistent system the computation cost of task $v_s$ on node $n_t$ is defined by $\omega(v_s, n_t)=r(v_s)/e(n_t)$, where $r(v_s)$ is the computation (or resource) requirement of task $v_s$, and $e(n_t)$ is the execution rate of node $n_t$. 

Let $T=\{t_1, t_2,\ldots, t_a\}$ be a set of $a$ tasks with or without precedence constraints and $N=\{n_1, n_2,\ldots, n_b\}$ be a set of $b$ nodes. Let $A:T \rightarrow N$ be a task assignment function between $T$ and $N$. Let $t_n^e(A)$ denote the total execution time of node $n$ for the task assignment $A$ and let $t_n^i(A)$ denote the total idle time of node $n$ for the task assignment $A$. The \emph{turnaround time} of node $n$ for the task assignment $A$ is the total time spent in the node $n$ for the task assignment $A$. Let $t_n (A) := t_n^i(A) + t_n^e(A)$ and $t(A) := \max_{n}t_n (A)$. We call $t(A)$ the \emph{task turnaround time} of the task assignment $A$. 

A \emph{2D mesh-connected system} (or \emph{2D-MS for short}) is a set of $m \times n$ nodes structured as a rectangular grid of height $m$ and width $n$. Each node is addressed by its coordinate $(i, j)$ for $1 \leq i \leq m$ and $1 \leq j \leq n$. An \emph{internal node} $(x, y)$, where $1 < x < m$ and $1 <y < n$, is directly connected to its four adjacent nodes $(x-1, y),\;(x+1, y),\;(x, y-1),\;(x, y+1)$. A node in the four corners has two adjacent nodes, while a node in the remaining boundary has three adjacent nodes, respectively (see Figure~\ref{fig:2DMeshTopologyGraphs}(d) in Section~\ref{sec:MeshGraphsAndYoungTableaux}). An $m^\prime \times n^\prime$ \emph{submesh} of an $m \times n$ 2D-MS is a grid of nodes belonging to the 2D-MS with height $m^\prime$ and width $n^\prime$ such that $1 \leq m^\prime \leq m$ and $1\leq n^\prime \leq n$. 
A submesh is called \emph{free} if every node in the submesh is idle. We assume that each incoming job (i.e., a set of tasks) requests a submesh of a certain size and that every node in the allocated submesh cannot be used for an incoming job until it is deallocated. We say that \emph{internal fragmentation} occurs if more nodes in a 2D-MS are allocated to a job than required. We say that \emph{external fragmentation} occurs if a large enough submesh cannot be found for an incoming job although there are a sufficient number of nodes in a 2D-MS are available.

A \emph{partition} of \emph{n} is defined as a sequence $\lambda=(\lambda_1, \lambda_2, \ldots, \lambda_i)$, where the $\lambda_k$ are weakly decreasing and $\sum_{k=1}^i{\lambda_k}=n$. If $\lambda$ is a partition of \emph{n}, then we write $\lambda \vdash n$.

Let $\lambda=(\lambda_1, \lambda_2, \ldots, \lambda_i) \vdash n$. A \emph{Young diagram} (or \emph{Ferrers diagram}) of shape $\lambda$ is a left-justified, finite collection of cells, with row \emph{j} containing $\lambda_j$ cells for $1 \leq j \leq i$. Each cell in a Young diagram in row $i$ and column $j$ has a coordinate $(i, j)$, as in a 2D-MS. 

Let $\lambda$ be a partition. An \emph{inner corner} of the Young diagram of shape $\lambda$ is a cell $(i, j) \in \lambda$ whose removal leaves the Young diagram of a partition. 

Let $\lambda \vdash n$. A \emph{tableau} \emph{t} of shape $\lambda$ is a Young diagram of shape $\lambda$ filled with a set of elements, often positive integers. An entry of a cell having a coordinate $(i, j)$ in tableau $t$ is denoted by $t_{i,j}$. 

A \emph{Young tableau} \emph{T} of shape $\lambda$ is a tableau of shape $\lambda$ whose entries are the numbers from 1 to $n$, each occurring once.

A \emph{standard Young tableau} is a Young tableau whose entries are strictly increasing in rows and columns. Let $\lambda \vdash n$. 

A \emph{partial tableau} $p$ of shape $\lambda$ is a tableau whose entries are strictly increasing in rows and columns. Note that a partial tableau is the standard Young tableau if the entries of $p$ are exactly $\{1,2, \ldots, n\}$. For instance, the following $t_1$ is a standard tableau, but $t_2$ is not. 
\begin{center}
$t_1=$ {$\young(135,24,6)$},\;\; $t_2=$ {$\young(135,42,6)$}.
\end{center}

The number of standard Young tableaux of a given shape $\lambda$ is obtained from the \emph{hook formula}.

If $\nu=(i, j)$ is a cell in the Young diagram of shape $\lambda$, then the \emph{hook} of $\nu$, denoted by $H_\nu$, is the set of all cells directly to the right of $\nu$ or directly below $\nu$ including $\nu$ itself, that is
\begin{center}
$H_\nu=H_{i,j}=\{(i,\,j^\prime):j^\prime \geq j\} \cup \{(i^\prime,\,j):i^\prime \geq i\}.$
\end{center}

The \emph{hook length} of $\nu=(i, j)$, denoted by $h_{i,j}$, is the number of cells in its hook, i.e., $h_{i,j}=|H_{i,j}|$.

The number of standard Young tableaux of a given shape $\lambda \vdash n$ is obtained by the following theorem.

\begin{thm}[\cite{frame1954}]
\label{thm:HookFormula}
If $\lambda \vdash n$, then the number $f^\lambda$ of standard Young tableaux of shape $\lambda$ is
\begin{center}
$f^\lambda=\dfrac{n!}{\prod_{(i,j) \in \lambda} h_{i,j}}$.
\end{center}
\end{thm}

For instance, labeling each cell with its hook length for the Young diagram of shape $(3, 2, 1)$ and $(4, 4, 4, 4)$ are given by
\begin{center}
$\young(531,31,1)$\;\;\;\;,\;\;\;\;$\young(7654,6543,5432,4321)$\;\;.
\end{center}
If the shape is $\lambda=(3, 2, 1)$, then $f^{(3, 2, 1)}=6!/(5\cdot3^2\cdot1^3)=16$. If the shape is $\lambda=(4, 4, 4, 4)$, then $f^{(4, 4, 4, 4)}=16!/(7\cdot6^2\cdot5^3\cdot4^4\cdot3^3\cdot2^2\cdot1)=24024$.

Let $\mu=(\mu_1, \mu_2,\ldots, \mu_i)$ and $\lambda=(\lambda_1, \lambda_2,\ldots, \lambda_j)$ be partitions with $\mu \subseteq \lambda$ (i.e., $i \leq j$ and $\mu_k \leq \lambda_k$ for $1 \leq k \leq i$). Then, a \emph{skew shape} of $\lambda/\mu$ is the set of cells $\lambda/\mu=\{c :c \in \lambda$ and $c \notin \mu\}$. A skew shape of $\lambda/\mu$ is \emph{normal} if $\mu=\emptyset$. A tableau of skew shape $\lambda / \mu$ is called a \emph{skew tableau} of shape $\lambda/\mu$. A \emph{partial skew tableau} of shape $\lambda/\mu$ (or a partial tableau of skew shape $\lambda/\mu$) is a skew tableau of shape $\lambda/\mu$ whose entries are strictly increasing in rows and columns. 

A partial skew tableau is called the \emph{standard skew tableau} if its entries are precisely $\{1, 2,\ldots, n\}$. For instance, consider skew tableaux of shape $\lambda/\mu=(4, 3, 3, 2)/(2, 2)$ given by
\begin{center}
$t_1 = $ {$\young(::17,::3,245,69)$}, \;\; $t_2 = $ {$\young(::17,::5,243,69)$} .
\end{center}
We see that $t_1$ is a partial skew tableau, but $t_2$ is not. 

A group $(G,\,\cdot\,)$ is a nonempty set \emph{G}, closed under a binary operation $\cdot$ , such that the following axioms are satisfied: (i) $(a\cdot b)\cdot c =  a \cdot (b \cdot c)$ for all $a,b,c \in G$, 
(ii) there is an identity element \emph{e} $\in$ \emph{G} such that for all $x\in G,~e \cdot x = x \cdot e = x$, (iii) for each element $a \in G$, there is an element $a^{-1} \in G$ such that $a \cdot a^{-1}= a^{-1} \cdot a = e$.

The group of all bijections $I_n  \rightarrow I_n$, whose binary operation is function composition, is called the \emph{symmetric group on n letters} and denoted $\mathfrak{S}_n$. Since $\mathfrak{S}_n$ is the group of all permutations of a set $I_n=\{1, 2,\ldots, n\}$, the order of $\mathfrak{S}_n$, i.e., $|\mathfrak{S}_n|$, is $n!$.

Let $i_1, i_2,\ldots, i_n$ be distinct elements of $I_n = \{1, 2,\ldots, n\}$. Then, $[i_1\,i_2\,\cdots\,i_n] \in \mathfrak{S}_n$ denotes the permutation that maps $1 \mapsto i_1,2 \mapsto i_2, \ldots, n \mapsto i_n$. 

Suppose $x < y <z$. A \emph{Knuth transformation} of a permutation $\pi \in \mathfrak{S}_n$ is a transformation of $\pi \in \mathfrak{S}_n$ into another permutation $\tau \in \mathfrak{S}_n$ that has one of the following forms:
\begin{my_enumerate}
\item $\pi = [x_1\, \cdots\, y\, x\, z\, \cdots\, x_n] \in \mathfrak{S}_n \Longrightarrow \tau = [x_1\, \cdots\, y\, z\, x\,  \cdots\, x_n] \in \mathfrak{S}_n$,
\label{enum:Knuthone}
\item $\pi = [x_1\, \cdots\, y\, z\, x\, \cdots\, x_n] \in \mathfrak{S}_n \Longrightarrow \tau = [x_1\, \cdots\, y\, x\, z\,  \cdots\, x_n] \in \mathfrak{S}_n$,
\label{enum:Knuthtwo}
\item $\pi = [x_1\, \cdots\, x\, z\, y\, \cdots\, x_n] \in \mathfrak{S}_n \Longrightarrow \tau = [x_1\, \cdots\, z\, x\, y\,  \cdots\, x_n] \in \mathfrak{S}_n$,
\label{enum:Knuththree}
\item $\pi = [x_1\, \cdots\, z\, x\, y\, \cdots\, x_n] \in \mathfrak{S}_n \Longrightarrow \tau = [x_1\, \cdots\, x\, z\, y\,  \cdots\, x_n] \in \mathfrak{S}_n$.
\label{enum:Knuthfour}
\label{defn:KnuthTransformationsfourth}
\end{my_enumerate}
Two permutations $\pi, \tau \in \mathfrak{S}_n$ are called \emph{Knuth-equivalent} if one of them can be obtained from the other by a sequence of Knuth transformations, denoted $\pi \cong_{K} \tau$. 

For instance, we see that $[2\,1\,3] \in \mathfrak{S}_3$ and $[2\,3\,1] \in \mathfrak{S}_3$ are Knuth-equivalent by the above \ref{enum:Knuthone} and \ref{enum:Knuthtwo}, written $[2\,1\,3] \cong_{K} [2\,3\,1]$. Similarly, $[1\,3\,2] \in \mathfrak{S}_3$ and $[3\,1\,2] \in \mathfrak{S}_3$ are Knuth-equivalent by the above \ref{enum:Knuththree} and \ref{enum:Knuthfour}, written $[1\,3\,2] \cong_{K} [3\,1\,2]$.

Let $t$ be a tableau. The \emph{reading word} or \emph{row word} of $t$, denoted $r(t)$, is the permutation of entries of $t$ obtained by concatenating the rows of $t$ from bottom to top, i.e., $r(t)=R_k R_{k-1} \ldots R_1$, where $R_1,\ldots, R_k$ are the rows of $t$.

\begin{algorithm}[h!]
\SetAlgoLined
\KwIn{A partial tableau $P$ of skew shape $\lambda/\mu$; an inner corner of $\mu$}
\KwOut{A partial tableau $P^\prime$}
\Begin
{
Pick $x$ to be an inner corner of $\mu$\;
\While{$x$ is not an inner corner of $\lambda$}
{
\If {$x=(i,j)$}
{
Let $x^\prime$ be the cell of $\text{min}\{P_{i+1,j}\,,\,P_{i, j+1}\}$\;(If only one of $P_{i+1,j}$ and $P_{i, j+1}$ exists, then choose that value as a minimum.)
}
Slide $P_{x^\prime}$ into cell $x$ and set $x := x^\prime$\;
}
\Return {The resulting partial tableau $P^\prime$}\;
}
\caption {A forward jeu de taquin slide~\cite{scutzenberger1963, sagan2001}}
\label{algorithm:JeuDeTaquinForwardSlides}
\end{algorithm}

The \emph{jeu de taquin} of Sc{\"u}tzenberger~\cite{scutzenberger1963, stanley1997} consists of a set of rules for transforming \emph{partial tableaux}, while some properties of partial tableaux are preserved during transformations. 
A forward jeu de taquin slide is described in Algorithm~\ref{algorithm:JeuDeTaquinForwardSlides}. Note that the resulting tableau of a forward jeu de taquin slide is still a partial tableau. We say that partial tableaux $P$ and $P^\prime$ are \emph{jeu de taquin equivalent}, written $P \cong_{jdt}P^\prime$, if $P^\prime$ can be obtained from $P$ by some sequence of jeu de taquin slides, or vice versa. (The reader is encouraged to verify that $\cong_{jdt}$ is an equivalence relation on the set of partial tableaux.)

\begin{lem}[\cite{stanley1997}]\label{lem:jdt}
Each jeu de taquin slide converts the reading word of a standard skew tableau into a Knuth-equivalent one.
\end{lem}
\begin{thm}[\cite{scutzenberger1963, sagan2001}]
\label{thm:jdt}
The jeu de taquin equivalence class of a given partial skew tableau $P$ contains exactly one partial tableau of normal shape.
\end{thm}

\begin{thm}[\cite{scutzenberger1963,stanley1997}]\label{thm:Kjdt}
Let $P$ and $P^\prime$ be standard skew tableaux. They are jeu de taquin equivalent, i.e., $P \cong_{jdt}P^\prime$, if and only if their reading words are Knuth-equivalent, i.e., $r(P) \cong_{K}r(P^\prime)$.
\end{thm}

\section{Representations of a 2D-HMS}
\label{sec:MeshGraphsAndYoungTableaux}

A graph-theoretic approach to Young diagrams or tableaux has already been researched in~\cite{Lee1989}. It focuses on graphs having the shape of a Young diagram or a tableau, while this paper focuses on converting a 2D-MS into a Young diagram. This section presents how a 2D-HMS is represented by a Young diagram with additional properties. In this section we define a hierarchical 2D mesh tableau in order to represent a task assignment and its reassignments in a 2D-HMS.

\begin{figure}[h!]
  \centering
      \includegraphics[width=0.95\textwidth]{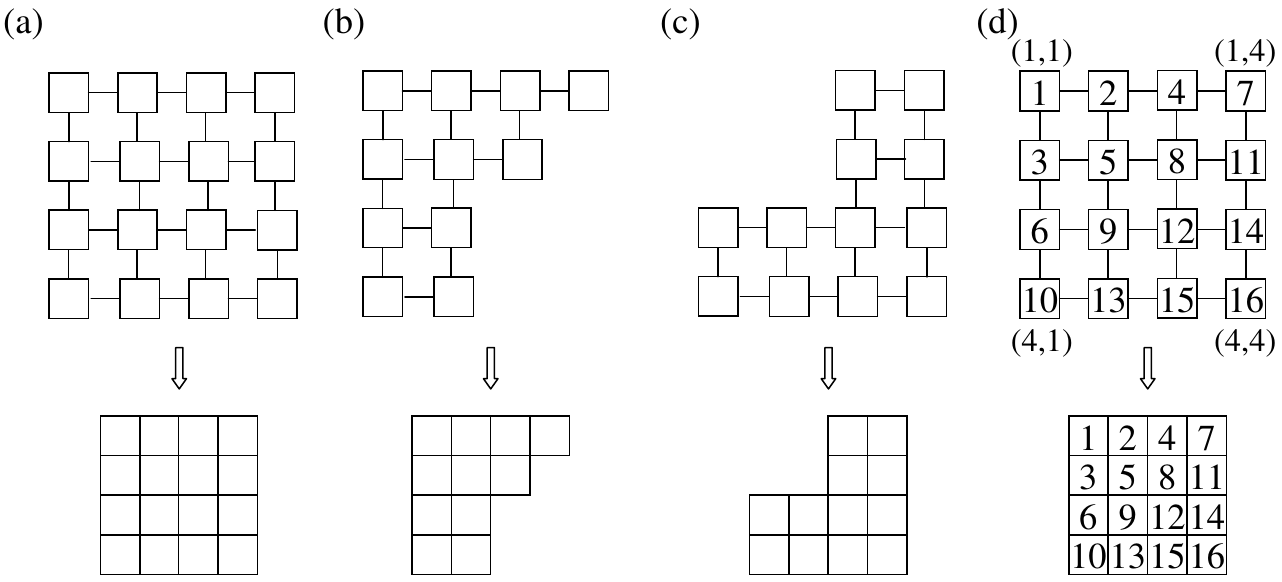}
 \caption{The conversion of each 2D-MS (or its variant) into a Young diagram or a tableau.}
\label{fig:2DMeshTopologyGraphs}
\end{figure}

Figure~\ref{fig:2DMeshTopologyGraphs}(a) shows our approach to convert a 2D-MS into a Young diagram in a compact manner. Similarly to Figure~\ref{fig:2DMeshTopologyGraphs}(a), Figure~\ref{fig:2DMeshTopologyGraphs}(b) and (c) convert the variants of a 2D-MS into their corresponding Young diagrams. Meanwhile, each label (except labels involving coordinates $(1, 1)$, $(1, 4)$, $(4, 1)$, and $(4, 4)$) in Figure~\ref{fig:2DMeshTopologyGraphs}(d) denotes a task ID in order to represent a task assignment in a 2D-MS.

To exploit the regular nature of a 2D mesh topology, we consider a hierarchical 2D mesh diagram of canonical shape, where the rows and columns of heterogeneous nodes are sorted in descending order by their priorities (or execution rates). We first define a 2D-HMS consisting of $m\times n$ heterogeneous nodes. Then, we define a hierarchical 2D mesh diagram to represent a 2D-HMS. 
\begin{defn}
A \emph{hierarchical 2D mesh-connected system} (2D-HMS) of $m \times n$ heterogeneous nodes is a heterogeneous 2D-MS with the following partial order
\begin{center}
$N(i-1, j) \prec_{p} N(i,j),\;\; N(i, j-1) \prec_{p} N(i, j),\;\;\; 1 <i \leq m,\; 1 < j \leq n$,
\end{center}
where $N(a, b) \prec_p N(c,d)$ means that a node addressed by $(a, b)$ has a higher priority (or execution rate) than a node addressed by $(c, d)$.
\end{defn}

\begin{defn}
\label{defn:HMD}
A \emph{hierarchical 2D mesh diagram of canonical shape} $\lambda=(\lambda_1, \lambda_2, \ldots, \lambda_k) \vdash n$ for $\lambda_1=\lambda_2=\cdots=\lambda_k$ is a Young diagram of shape $\lambda \vdash n$, where each cell $(i, j)$ represents each node $(i, j)$ in a 2D-HMS. Therefore, it has the following partial order
\begin{center}
$N(i-1, j) \prec_p N(i,j),\;\; N(i, j-1) \prec_p N(i, j),\;\;\; 1 <i \leq k,\; 1 < j \leq \lambda_1$,
\end{center}
where $N(a, b) \prec_p N(c,d)$ means that a node represented by cell $(a, b)$ has a higher priority (or execution rate) than a node represented by cell $(c, d)$.
\end{defn}

We say ``a node represented by cell $(a, b)$'' in Definition~\ref{defn:HMD} and ``a node addressed by $(a, b)$'' interchangeably for a hierarchical 2D mesh diagram. The following proposition involves in a counting aspect of organizing a hierarchical 2D mesh diagram of canonical shape $\lambda \vdash n$ using $n$ heterogeneous nodes in a distributed system.
\begin{prop}
Let $\lambda=(\lambda_1, \lambda_2, \ldots, \lambda_k) \vdash n$ for $\lambda_1=\lambda_2=\cdots=\lambda_k$ and let $N=\{1, 2,\ldots, n\}$ be a set of $k \times \lambda_1$ heterogeneous nodes having priorities\footnote{In some priority schemes~\cite{kwok1999, sinnen2007} a higher number indicates a higher priority. Throughout this paper it is assumed that a lower number indicates a higher priority.} represented by node IDs from 1 to $n$. A total order relation $<$ is defined naturally on $N$ such that for any two nodes $u \in N$ and $v \in N$, $u<v$ implies that node $u$ has a higher priority (or execution rate) than node $v$. Then, the number of ways to organize a hierarchical 2D mesh diagram of shape $\lambda \vdash n$ by arranging $n$ nodes in $N$ is $f^\lambda$ (i.e., the number of standard Young tableaux of shape $\lambda$).
\end{prop}
\begin{proof}
It immediately follows from the definition of a standard tableau and Definition~\ref{defn:HMD}.
\end{proof}

We next define a \emph{hierarchical 2D mesh tableau} whose main usage is to represent a priority-based task assignment and its reassignments in a 2D-HMS. A priority-based task assignment and its reassignments in a 2D-HMS are discussed in the next section.

\begin{defn}
\label{defn:HMT}
A \emph{hierarchical 2D mesh tableau of shape} $\lambda \vdash n$ is a tableau of shape $\lambda$ whose underlying Young diagram is a hierarchical 2D mesh diagram of canonical shape $\lambda \vdash n$. It is denoted by $(\text{HMT}_{i,j})$ of shape $\lambda$ or $\lambda$-$(\text{HMT}_{i,j})$ for short. Entries of $\lambda$-$(\text{HMT}_{i,j})$ denote task IDs from the set $\{1, 2,\ldots, m\}$ for $m\leq n$. The empty entry of a cell is allowed in $\lambda$-$(\text{HMT}_{i,j})$, while all non-empty entries along with their cells must form a tableau of normal or skew shape, called a \emph{maximally embedded tableau} of $\lambda$-$(\text{HMT}_{i,j})$.
\end{defn}

\begin{defn}
Let $\lambda$-$(\text{HMT}_{i,j})$ be a hierarchical 2D mesh tableau of shape $\lambda \vdash n$. If the maximally embedded tableau of $\lambda$-$(\text{HMT}_{i,j})$ is a tableau of normal shape, we say that $\lambda$-$(\text{HMT}_{i,j})$ is of \emph{normal shape}. Meanwhile, if the maximally embedded tableau of $\lambda$-$(\text{HMT}_{i,j})$ is a tableau of skew shape, we say that $\lambda$-$(\text{HMT}_{i,j})$ is of \emph{skew shape}. If the maximally embedded tableau of $\lambda$-$(\text{HMT}_{i,j})$ is a partial tableau (i.e., a tableau whose entries are strictly increasing in rows and columns), we say that $\lambda$-$(\text{HMT}_{i,j})$ is \emph{standard}. Otherwise, we say that $\lambda$-$(\text{HMT}_{i,j})$ is \emph{generalized}.
\end{defn}



\section{Priority-based task reassignments in a 2D-HMS}
\label{sec:Priority-basedTaskReassignmentsAndYoungTableaux}

Let $T_m=\{1, 2,\ldots, m\}$ be a set of $m$ tasks having priorities represented by task IDs from 1 to $m$, where a lower task ID indicates a higher priority. A total order relation $\prec_t$ is defined on $T_m$ as follows. $t_1 \prec_t t_2$ for any two tasks $t_1\in T_m$ and $t_2 \in T_m$ means that $t_1$ has a higher priority than $t_2$. Let $R_n$ be a hierarchical 2D-MS (2D-HMS) consisting of $n\,(n \geq m)$ heterogeneous nodes whose priorities (or execution rates) of rows and columns are sorted in descending order. Let $A:T_m \rightarrow R_n$ be an injective task assignment function between $T_m$ and $R_n$ whose constraints are defined as follows:
\begin{enumerate}
\item Priorities of the assigned tasks on nodes strictly decrease in rows and columns, i.e., $t_1 \prec_t t_2$ whenever $A(t_1) \prec_p A(t_2)$ for any two tasks $t_1 \in T_m$ and $t_2 \in T_m$.
\item Nodes of $A(T_m)$ in $R_n$ is left-justified, where the row sizes of $A(T_m)$ are weakly decreasing.
\end{enumerate}

The first constraint ensures that a node with a higher priority executes a task with a higher priority. The second constraint ensures that if one node is idle and the other node is busy for two adjacent nodes in a 2D-HMS, the node with the lower priority is chosen to be idle. We say that a task assignment or reassignment $A$ in a 2D-HMS is \emph{priority-based} if it satisfies the above constraints in a 2D-HMS. 

The \emph{priority-based task reassignment problem} in a 2D-HMS is defined as follows: We are given an initial task assignment $A=A_0$ along with a task completion sequence $(b_i)_{i=1}^{m}$ provided in run-time, where $b_i \in T_m$ for $1 \leq i \leq m$. Find a priority-based task (re)assignment sequence $(A_k)_{k=0}^{m-1}$. 

The constraints and assumptions that we have made are:
\begin{enumerate}
\item Both tasks and nodes are heterogeneous.
\item Each node can process at most one task at a time.
\item Each priority-based task reassignment is achieved by an iterative sequence of task relocations, where each task relocation is allowed between two adjacent nodes in a 2D-HMS if one node is in idle state and the other node is in busy state.  
\item Two task relocations do not occur simultaneously in a 2D-HMS.
\item The number of tasks are less than or equal to the number of the available nodes in a 2D-HMS. 
\end{enumerate}
Recall that the underlying Young diagram of $\lambda$-$(\text{HMT}_{i,j})$ is a hierarchical 2D mesh diagram of canonical shape (see Definition~\ref{defn:HMT}). Therefore, each cell of $\lambda$-$(\text{HMT}_{i,j})$ represents each node in a 2D-HMS. By assigning each entry (i.e., task) in a standard $\lambda$-$(\text{HMT}_{i,j})$ of normal shape to its underlying cell representing a node in a 2D-HMS, we see that a standard $\lambda$-$(\text{HMT}_{i,j})$ of normal shape represents a priority-based task assignment in a 2D-HMS. We define a \emph{descent pair} of a generalized $\lambda$-$(\text{HMT}_{i,j})$, which does not allow a generalized $\lambda$-$(\text{HMT}_{i,j})$ to represent a priority-based task assignment in a 2D-HMS.

\begin{defn}
Let $\lambda$-$(\text{HMT}_{i,j})$ be a generalized hierarchical 2D mesh tableau. 
If $\text{HMT}_{i,j}\prec_t\text{HMT}_{i-1,j}$ (respectively, $\text{HMT}_{i,j}\prec_t\text{HMT}_{i,j-1}$), then, $\{(i-1, j), (i, j)\}$ (respectively, $\{(i, j-1), (i, j)\}$) is called a \emph{descent pair} of a generalized $\lambda$-$(\text{HMT}_{i,j})$.
\end{defn}

If a descent pair occurs in a task assignment represented by a generalized $\lambda$-$(\text{HMT}_{i,j})$, it is not a priority-based task assignment. In this paper $\lambda$-$(\text{HMT}_{i,j})$ is referred to as a standard $\lambda$-$(\text{HMT}_{i,j})$ of normal shape unless otherwise stated.

Now, consider a priority-based task assignment in Figure~\ref{fig:TaskReassignments}(a) represented by $(\text{HMT}_{i,j})$ of shape $\lambda=(3,3,3) \vdash n$ for $n=9$. 

\begin{figure}[h!]
(a)\;$A_0=\young(124,357,689)$\;,\;\;(b)\;$\young(\bullet24,357,689)\Rightarrow\young(2\bullet4,357,689)\Rightarrow\young(24\bullet,357,689)\Rightarrow \young(247,35\bullet,689)\Rightarrow\young(247,359,68\bullet)$\;.
\linebreak \linebreak

(c)\;$A_0=\young(124,357,689)$\;,\;$A_1=\young(247,359,68\hfil)$\;,\;$A_2=\young(247,589,6\hfil\hfil)$\;,\;$A_3=\young(479,58\hfil,6\hfil\hfil)$\;,\;$A_4=\young(479,68\hfil,\hfil\hfil\hfil)$\;,
\begin{center}
$A_5=\young(479,6\hfil\hfil,\hfil\hfil\hfil)$\;,\;$A_6=\young(679,\hfil\hfil\hfil,\hfil\hfil\hfil)$\;,\;$A_7=\young(79\hfil,\hfil\hfil\hfil,\hfil\hfil\hfil)$\;,\;$A_8=\young(9\hfil\hfil,\hfil\hfil\hfil,\hfil\hfil\hfil)$\;\;.
\end{center}
  \caption{A sequence of task reassignments for a task completion sequence (1,\,3,\,2,\,5,\,8,\,4,\,6,\,7,\,9).}
\label{fig:TaskReassignments}
\end{figure}

\begin{algorithm}[h!]
\SetAlgoLined
\KwIn{An initial task assignment $A_0$ represented by $\lambda$-$(\text{HMT}_{i,j})$ of normal shape; a task completion sequence $(b_1, b_2,\ldots, b_{m})$, where $m \geq 2$, provided in runtime}
\KwOut{A task (re)assignment sequence $(A_k)_{k=0}^{m-1}$}
\Begin
{
Set $\lambda$-$(\text{HMT}^{(1)}_{i,j})$ :=$\lambda$-$(\text{HMT}_{i,j})$\;
\For{$k\leftarrow 1$ \KwTo $m-1$}
{
Let $t$ be the maximally embedded tableau of $\lambda$-$(\text{HMT}^{(k)}_{i,j})$ and $\mu$ be the shape of $t$; Wait for a completion of task $b_k$ in $\lambda$-$(\text{HMT}^{(k)}_{i,j})$; If task $b_k$ is completed, then the underlying cell of task $b_k$ in $\lambda$-$(\text{HMT}^{(k)}_{i,j})$ becomes vacated; Set $x$ as the corresponding cell in $\lambda$-$(\text{HMT}^{(k)}_{i,j})$\;
\While{$x$ is not an inner corner of $\mu$}
{
\If {$x=(i,\,j)$}
{
Let $x^\prime$ be the cell of $\text{min}\{\text{HMT}^{(k)}_{i+1,j}\,,\,\text{HMT}^{(k)}_{i, j+1}\}$. ( If only one non-idle cell exists in $(i+1,j)$ and $(i, j+1)$, then choose that cell.)
}
Relocate the task on cell $x^\prime$ into cell $x$ and set $x := x^\prime$\;
}
Set $A_k$:=$\lambda$-$(\text{HMT}^{(k)}_{i,j})$, where $\lambda$-$(\text{HMT}^{(k)}_{i,j})$ is of normal shape\;
Set $\lambda$-$(\text{HMT}^{(k+1)}_{i,j})$:=$\lambda$-$(\text{HMT}^{(k)}_{i,j})$\;
}
\Return {The task (re)assignment sequence $(A_k)_{k=0}^{m-1}$}\;
}
\caption {Task reassignments: $\lambda$-$(\text{HMT}_{i,j})$ of normal shape}
\label{algorithm:NormalTaskReassignments}
\end{algorithm}
If task 1 in Figure~\ref{fig:TaskReassignments}(a) is completed first, the node addressed by $(1, 1)$ becomes idle. We mark the cell $(1,1)$ as $\bullet$ to show that the node addressed by $(1, 1)$ is now in the idle state. Once a node is in the idle state, it seeks the right and below node to perform task relocation. Recall that our task relocation is only allowed between two adjacent nodes if one is in the idle state and the other is in the busy state. If a node is in the idle state, it does not check the left and above node to perform task relocation. It is because the underlying 2D mesh Young diagram of $\lambda$-$(\text{HMT}_{i,j})$ is hierarchical, it is not an optimal choice if a task is to run on a node with the lower execution rate. Therefore, a node in the idle state always seeks both the right and below node in order to compare task priorities and to relocate a task. We see that task $\text{HMT}_{1,2}$ has a higher priority than task $\text{HMT}_{2,1}$, i.e., $2 \prec_t 3$. Therefore, task relocation involves the node addressed by $(1,1)$ and the node addressed by $(1,2)$. Now, task 2 has been relocated and the node addressed by $(1,2)$ becomes idle. If a node becomes idle, the choice for task relocation between the right and below node is always \emph{greedy}~\cite{cormen2001}, allowing the task with the higher priority to occupy the idle node (see Figure~\ref{fig:TaskReassignments}(b)). This process continues until no task relocation is possible, which means that the right and below node of $\bullet$ are both idle or both not available. The final state of Figure~\ref{fig:TaskReassignments}(b) is the task reassignment $A_1$ for the completion of task 1. Given an initial task assignment $A_0$ in Figure~\ref{fig:TaskReassignments}(a), Figure~\ref{fig:TaskReassignments}(c) shows the task (re)assignment sequence $(A_k)_{k=0}^{m-1}$ for $m=9$ corresponding to the task completion sequence $(1,\,3,\,2,\,5,\,8,\,4,\,6,\,7,\,9)$. Algorithm~\ref{algorithm:NormalTaskReassignments} describes the procedure in Figure~\ref{fig:TaskReassignments}. Task reassignments take place in Algorithm~\ref{algorithm:NormalTaskReassignments} when a task completion sequence is provided in run time. It turns out that the iterative greedy task relocation mechanism in Algorithm~\ref{algorithm:NormalTaskReassignments} corresponds to a forward jeu de taquin slide discussed in Algorithm~\ref{algorithm:JeuDeTaquinForwardSlides}, except that task relocation starts with the cell that is indicated by the task completion sequence. Note that each task assignment in a task (re)assignment sequence $(A_k)_{k=0}^{m-1}$ in Algorithm~\ref{algorithm:NormalTaskReassignments} is a priority-based task assignment. We see that each task (re)assignment in $(A_k)_{k=0}^{m-1}$ in Algorithm~\ref{algorithm:NormalTaskReassignments} is represented by a hierarchical 2D mesh tableau of normal shape, where task IDs are increasing in rows and columns on the underlying hierarchical 2D mesh diagram. 

Thus far, we have examined the case where an initial task assignment is represented by $\lambda$-$(\text{HMT}_{i,j})$ of normal shape. We now consider the case where an initial task assignment is represented by  $\lambda$-$(\text{HMT}_{i,j})$ of skew shape.
\begin{figure}[h!]
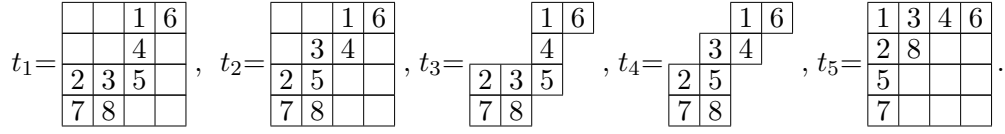

$t_1$=\,{$\young(\hfil\hfil16,\hfil\hfil4\hfil,235\hfil,78\hfil\hfil)$}\;,\; $t_2$=\,{$\young(\hfil\hfil16,\hfil34\hfil,25\hfil\hfil,78\hfil\hfil)$}\;,\;$t_3$=\,{$\young(::16,::4,235,78)$}\;,\;$t_4$=\,{$\young(::16,:34,25,78)$}\;,\;$t_5$=\,{$\young(1346,28\hfil\hfil,5\hfil\hfil\hfil,7\hfil\hfil\hfil)$}\,.
\caption{Task reassignments by using forward jeu de taquin slides.}
\label{fig:NormalReassignments}
\end{figure}

\begin{algorithm}[h!]
\SetAlgoLined
\KwIn{An initial task assignment $A_0$ represented by $\lambda$-$(\text{HMT}_{i,j})$ of skew shape}
\KwOut{A task (re)assignment sequence $(A_k)_{k=0}^{m}$}
\Begin
{
If $\beta$ is a partition of a positive integer $n$ for the maximally embedded tableau of $\lambda$-$(\text{HMT}_{i,j})$ of skew shape $\alpha/\beta$, then set $m$ as the value of $n$;
Set $\lambda$-$(\text{HMT}^{(1)}_{i,j})$ :=$\lambda$-$(\text{HMT}_{i,j})$; Set $k:=1$\;
\While{the maximally embedded tableau of $\lambda$-$(\text{HMT}^{(k)}_{i,j})$ is not of normal shape}
{
If the maximally embedded tableau of $\lambda$-$(\text{HMT}^{(k)}_{i,j})$ is of (skew) shape $\mu/\nu$, pick $x$ to be an inner corner of $\nu$\;
\While{$x$ is not an inner corner of $\mu$}
{
\If {$x=(i,\,j)$}
{
Let $x^\prime$ be the cell of $\text{min}\{\text{HMT}^{(k)}_{i+1,j}\,,\,\text{HMT}^{(k)}_{i, j+1}\}$. (If only one non-idle cell exists in $(i+1,j)$ and $(i, j+1)$, then choose that cell.)
}
Relocate the task on cell $x^\prime$ into cell $x$ and set $x := x^\prime$\;
}
Set $A_k$:=$\lambda$-$(\text{HMT}^{(k)}_{i,j})$; 
Set $\lambda$-$(\text{HMT}^{(k+1)}_{i,j})$:=$\lambda$-$(\text{HMT}^{(k)}_{i,j})$; 
$k:=k+1$\;
}
\Return {
The task (re)assignment sequence $(A_k)_{k=0}^{m}$}, where $A_{m}$ is the task reassignment represented by $\lambda$-$(\text{HMT}^{(m)}_{i,j})$ of normal shape\;
}
\caption{Task reassignments: $\lambda$-$(\text{HMT}_{i,j})$ of skew shape}
\label{algorithm:SkewedTaskReassignments}
\end{algorithm}

Consider a top-left corner of a hierarchical 2D mesh tableau $t_1$ or $t_2$ in Figure~\ref{fig:NormalReassignments}, where the node with the highest priority is idle. Therefore, it is a natural choice to relocate a task from a node with the lower execution rate to a node with the higher execution rate if task relocation is necessary. Algorithm~\ref{algorithm:SkewedTaskReassignments} describes the procedure, where an initial task assignment represented by a hierarchical 2D mesh tableau of skew shape is converted into the task reassignment represented by a hierarchical 2D mesh tableau of normal shape. As shown in Figure~\ref{fig:NormalReassignments}, $t_3$ is the maximally embedded tableau of $t_1$, and $t_4$ is the maximally embedded tableau of $t_2$, respectively. By performing task relocations discussed in Algorithm~\ref{algorithm:SkewedTaskReassignments} iteratively, the task reassignment $t_5$ is obtained from the task assignment $t_1$ in Figure~\ref{fig:NormalReassignments} by Algorithm~\ref{algorithm:SkewedTaskReassignments}. Similarly, the task reassignment $t_5$ is also obtained from the task assignment $t_2$ in Figure~\ref{fig:NormalReassignments}. If $t_2$ represents an initial task assignment $A_0$, then $t_5$ corresponds to the task reassignment $A_3$ in Algorithm~\ref{algorithm:SkewedTaskReassignments}. An initial choice of task relocation for $A_1$ must target for either the cell $(1, 2)$ or the cell $(2, 1)$ in $t_2$. Note that the task (re)assignment sequence $(A_0, A_1, A_2, A_3)$ is not uniquely determined, depending on the choice of an initial task relocation. However, $A_3$ is uniquely determined by Theorem~\ref{thm:jdt} because the greedy-based task relocations on $\lambda$-$(\text{HMT}_{i,j})$ follow the forward jeu de taquin slide rules. Unlike a task reassignment represented by $\lambda$-$(\text{HMT}_{i,j})$ of normal shape, a task reassignment represented by $\lambda$-$(\text{HMT}_{i,j})$ of skew shape do not always satisfy the constraints of a priority-based task assignment. For instance, non-idle nodes of $A_0$, $A_1$, and $A_2$ are not left-justified, which implies that the node with the higher execution rate is idle for some pairs of adjacent nodes in a 2D-HMS. However, the resulting task reassignment $A_3$ is a priority-based task assignment in a 2D-HMS. We next define an equivalence class of task reassignments up to task relocations under the greedy task relocation policy.

\begin{defn}
\label{defn:TaskReassignmentEquivalent}
Two task assignments $t_1$ and $t_2$, represented by a hierarchical 2D mesh tableau $\lambda$-$(\text{HMT}^{1}_{i,j})$ of skew shape and $\lambda$-$(\text{HMT}^{2}_{i,j})$ of skew shape, respectively, are called \emph{task reassignment equivalent} up to task relocations (under the greedy task relocation policy described in Algorithm~\ref{algorithm:SkewedTaskReassignments}), denoted by $t_1 \cong_{t} t_2$, if they have the same resulting task reassignment represented by the same $\lambda$-$(\text{HMT}_{i,j})$ of normal shape.
\end{defn}

\begin{prop}
Let $\lambda$-$(\text{HMT}^{1}_{i,j})$ and $\lambda$-$(\text{HMT}^{2}_{i,j})$ be hierarchical 2D mesh tableaux of skew shape whose maximally embedded tableaux are both standard skew tableaux. If two task assignments $t_1$ and $t_2$, represented by $\lambda$-$(\text{HMT}^{1}_{i,j})$ and $\lambda$-$(\text{HMT}^{2}_{i,j})$, respectively, are task reassignment equivalent, then the reading words of their maximally embedded tableaux of $\lambda$-$(\text{HMT}^{1}_{i,j})$ and $\lambda$-$(\text{HMT}^{2}_{i,j})$ are Knuth-equivalent.
\end{prop}
\begin{proof}
Let $P$ be the maximally embedded tableau of $\lambda$-$(\text{HMT}^{1}_{i,j})$ for the task assignment $t_1$, and $Q$ be the maximally embedded tableau of $\lambda$-$(\text{HMT}^{2}_{i,j})$ for the task assignment $t_2$. Since $t_1 \cong_{t} t_2$ by hypothesis, $t_1$ and $t_2$ have the same resulting task reassignment represented by the same  $\lambda$-$(\text{HMT}_{i,j})$ of normal shape by Definition~\ref{defn:TaskReassignmentEquivalent}. Each task relocation under the greedy task relocation policy described in Algorithm~\ref{algorithm:SkewedTaskReassignments} follows the forward jeu de taquin slide rules described in Algorithm~\ref{algorithm:JeuDeTaquinForwardSlides}. Thus, $P \cong_{jdt}Q$ by Theorem~\ref{thm:jdt}. Since $P$ and $Q$ are both standard skew tableaux by hypothesis satisfying $P \cong_{jdt}Q$, we conclude that the reading words of $P$ and $Q$ are Knuth-equivalent by Theorem~\ref{thm:Kjdt}.
\end{proof}

For instance, the reading word of $t_3$ (i.e., the maximally embedded tableau of $t_1$) in Figure~\ref{fig:NormalReassignments} is $\pi=[7\,8\,2\,3\,5\,4\,1\,6] \in \mathfrak{S}_8$. Similarly, the reading word of $t_4$ (i.e., the maximally embedded tableau of $t_2$) in Figure~\ref{fig:NormalReassignments} is $\tau=[7\,8\,2\,5\,3\,4\,1\,6] \in \mathfrak{S}_8$. We leave it to the reader to verify that $\pi$ and $\tau$ are Knuth-equivalent.

We say that an idle node in a 2D-HMS represented by $\lambda$-$(\text{HMT}_{i,j})$ is \emph{locally fragmented} if it is surrounded by busy nodes, where a busy node in a 2D-HMS corresponds to a cell having an non-empty entry in $\lambda$-$(\text{HMT}_{i,j})$. If we do not apply a task reassignment at all, a locally fragmented node can be generated in a 2D-HMS when an internal node completes its task and becomes idle while other nodes are busy. It is easy to see that locally fragmented nodes increase the chance of occurring external fragmentation. For instance, four locally fragmented nodes can serve only four $1\times1$ meshes of tasks, but cannot serve one $2\times2$ mesh of tasks or two $1\times2$ meshes of tasks. As discussed in~\cite{yoo2000}, task relocation is an approach for alleviating the fragmentation problem in a 2D-MS. In our approach we use task relocations to avoid the generation of locally fragmented nodes while taking the priorities of tasks and nodes into account. The following proposition says that our priority-based task reassignment procedure does not generate any locally fragmented node in a 2D-HMS. 

\begin{prop}
Let $A_0$ be an initial task assignment in a 2D-HMS represented by $\lambda$-$(\text{HMT}_{i,j})$ of normal shape and let $(b_1, b_2,\ldots, b_m)$ for $m\geq 2$ be a task completion sequence in Algorithm~\ref{algorithm:NormalTaskReassignments}. Each task reassignment $A_k$ ($k=1,\ldots,m-1$) in Algorithm~\ref{algorithm:NormalTaskReassignments} does not generate any locally fragmented node in the 2D-HMS.
\end{prop}
\begin{proof}
For each $k=0,\ldots,m-1$, let $t_k$ be the maximally embedded tableau of $\lambda$-$(\text{HMT}^{k}_{i,j})$ representing task (re)assignment $A_k$ in a 2D-HMS and let $i_k$ be the inner corner in $t_k$ that becomes vacated by $A_{k+1}$ by Algorithm~\ref{algorithm:NormalTaskReassignments}. We see that $t_k$ is a tableau of normal shape by Algorithm~\ref{algorithm:NormalTaskReassignments}. By the definition of an inner corner, each inner corner in $t_k$ is adjacent to a cell with the empty entry (i.e., an idle node in the 2D-HMS) in $\lambda$-$(\text{HMT}^{k}_{i,j})$. Since the node addressed by $i_k$ does not become locally fragmented by $A_{k+1}$, we see that each task reassignment $A_k$ ($k=1,\ldots,m-1$) in Algorithm~\ref{algorithm:NormalTaskReassignments} does not generate any locally fragmented node in the 2D-HMS.
\end{proof}

As a simple example of Algorithm~\ref{algorithm:NormalTaskReassignments}, consider a sequential task assignment in a 2D-HMS for priority-based task reassignments, which is described as follows. A set of $m$ tasks $T_m=\{1, 2,\ldots, m\}$ with the precedence relationship $1 \rightarrow 2 \rightarrow\cdots\rightarrow m$ are to be assigned to a set of $m$ heterogeneous nodes bijectively in a 2D-HMS of $n$ heterogeneous nodes $(m \leq n)$ and executed sequentially without gaps. We assume that a 2D-HMS is consistent for a sequential task assignment. Tasks are heterogeneous, and their priorities are assigned by their computation requirements in which the higher task priority indicates the larger computation requirement. If a descent pair occurs in a sequential task assignment represented by a generalized $\lambda$-$(\text{HMT}_{i,j})$, there is a node with a higher priority than the other node but it executes a task with a lower priority than the other node. Therefore, it always has the better task assignment in terms of task turnaround time. For instance, swapping tasks on a descent pair reduces task turnaround time for a sequential task assignment in a consistent 2D-HMS. Now, consider priority-based task reassignments in Algorithm~\ref{algorithm:NormalTaskReassignments} for a priority-based sequential task assignment of $m$ tasks in a 2D-HMS, in which the task completion sequence is simply $(1, 2,\ldots, m)$. If we assume that every 2D-HMS is consistent in which task relocations are cost-free, we have Proposition~\ref{prop:SequentialSchedulingTaskReassignments}.

\begin{prop}
\label{prop:SequentialSchedulingTaskReassignments}
Let $A_0$ be a priority-based sequential task assignment for $m\;(m \geq 2)$ tasks represented by  $\lambda$-$(\text{HMT}_{i,j})$ of normal shape. Let $T_1$ be the task turnaround time for $A_0$ without task relocation, and $T_2$ be the task turnaround time with the task (re)assignment sequence $(A_k)_{k=0}^{m-1}$ (see Algorithm~\ref{algorithm:NormalTaskReassignments}). Then, $T_2$ is less than $T_1$, i.e., $T_2 < T_1$.
\end{prop}

\begin{proof}
It suffices to show that each task relocation allows each task to reduce its task execution time for a sequential task assignment in a 2D-HMS. Let $t_{i,x}$ be the task execution time of task $i$ on cell $x$ in $\lambda$-$(\text{HMT}_{i,j})$ before task relocation. After task relocation, task $i$ is relocated from cell $x$ to its adjacent cell $x^\prime$ such that $N(x^\prime) \prec_p N(x)$, where $N(x^\prime) \prec_p N(x)$ means that an execution rate of node addressed by $x^\prime$ is higher than that of node addressed by $x$. Thus, $t_{i,x^\prime} < t_{i,x}$. Since each task reassignment consists of iterative task relocations by Algorithm~\ref{algorithm:NormalTaskReassignments}, we conclude that $T_2 < T_1$.
\end{proof}

However, when we consider a non-trivial task relocation cost in a consistent 2D-HMS, the sufficient condition for task $i$ in a priority-based sequential task assignment to reduce its task turnaround time by means of task relocation from node $n$ to its adjacent node $n^\prime$ in Algorithm~\ref{algorithm:NormalTaskReassignments} is that the task relocation cost of task $i$ from node $n$ to $n^\prime$ is less than the difference of task execution times caused by task relocation of task $i$ from node $n$ to $n^\prime$. 
Note that a greedy task relocation policy in Algorithm~\ref{algorithm:NormalTaskReassignments} keeps $\lambda$-$(\text{HMT}_{i,j})$ from occurring any descent pair for each task reassignment.

Given a priority-based sequential task assignment in a 2D-HMS without any descent pair, we have discussed priority-based task reassignments in a 2D-HMS using Algorithm~\ref{algorithm:NormalTaskReassignments}. We leave it as an open question to apply the priority-based task reassignment procedure in a 2D-HMS to other application areas, such as sorting on a mesh-connected computing environment~\cite{thompson1977}.

\section{Conclusions}
\label{sec:ConclusionsOfPrioritybasedTaskAssignment}
In this paper we have presented a novel approach to representing priority-based task reassignments in a heterogeneous 2D mesh-connected system. To the best of our knowledge, this paper is the first attempt to study task assignments and reassignments in a 2D mesh-connected system using tableaux. We have proposed a hierarchical 2D mesh-connected system (2D-HMS) that is a heterogeneous 2D-MS in a distributed system with additional priority constraints on nodes. A priority-based task reassignment in a 2D-HMS is represented by a hierarchical 2D mesh tableau $\lambda$-$(\text{HMT}_{i,j})$ in which task relocations under the greedy task relocation policy are reduced to a jeu de taquin slide on $\lambda$-$(\text{HMT}_{i,j})$. Given an initial priority-based task assignment in a 2D-HMS represented by $\lambda$-$(\text{HMT}_{i,j})$, we have shown that our task reassignment procedure does not generate any locally fragmented node in the 2D-HMS while taking priorities of tasks and nodes into account.

\nocite{*}
\footnotesize
\begin{spacing}{0.1}
\bibliographystyle{plain}
\bibliography{dkim}
\end{spacing}
\end{document}